\documentclass[prl,aps,showpacs,twocolumn,superscriptaddress,amsmath,amssymb]{revtex4}

\usepackage{multirow}
\usepackage{graphicx,dcolumn,bm}
\usepackage{mathrsfs}
\usepackage[usenames,dvipsnames]{color}
\usepackage[colorlinks=true,citecolor=MidnightBlue,linkcolor=MidnightBlue,urlcolor=MidnightBlue]{hyperref}

\newcommand{\average}[1]{\langle #1 \rangle}

\newcommand{\fr}{\mathbf{r}}

\newcommand{\E}[1]{\hat{E}^{( #1 )}}

\newcommand{\ket}[1]{|#1\rangle}

\renewcommand{\vec}[1]{{\bf #1}}
\newcommand {\staf}{\begin{equation}}
\newcommand {\stof}{\end{equation}}
\newcommand {\staffeld}{\begin{eqnarray}}
\newcommand {\stoffeld}{\end{eqnarray}}

\begin{document}

\title{Directional superradiant emission from statistically independent incoherent non-classical and classical sources}

\author{S.~Oppel}
\affiliation{Institut f\"{u}r Optik, Information und Photonik, Universit\"{a}t Erlangen-N\"{u}rnberg, Erlangen, Germany}
\affiliation{Erlangen Graduate School in Advanced Optical Technologies (SAOT), Universit\"at Erlangen-N\"urnberg, Erlangen, Germany}

\author{R.~Wiegner}
\affiliation{Institut f\"{u}r Optik, Information und Photonik, Universit\"{a}t Erlangen-N\"{u}rnberg, Erlangen, Germany}

\author{G.~S.~Agarwal}
\affiliation{Erlangen Graduate School in Advanced Optical Technologies (SAOT), Universit\"at Erlangen-N\"urnberg, Erlangen, Germany}
\affiliation{Department of Physics, Oklahoma State University, Stillwater, OK, USA}

\author{J.~von~Zanthier}
\affiliation{Institut f\"{u}r Optik, Information und Photonik, Universit\"{a}t Erlangen-N\"{u}rnberg, Erlangen, Germany}
\affiliation{Erlangen Graduate School in Advanced Optical Technologies (SAOT), Universit\"at Erlangen-N\"urnberg, Erlangen, Germany}

\date{\today}

\begin{abstract}
Superradiance is one of the outstanding problems in quantum optics since Dicke introduced the concept of enhanced directional spontaneous emission by an ensemble of identical two-level atoms. The effect is based on correlated collective Dicke states which turn out to be highly entangled. Here we show that enhanced directional emission of spontaneous radiation can be produced also with statistically independent incoherent sources via the measurement of higher order correlation functions of the emitted radiation. Our analysis is applicable to a wide variety of quantum systems like trapped atoms, ions, quantum dots or NV-centers, and is also valid for statistically independent incoherent classical emitters. This is experimentally confirmed with up to eight independent thermal light sources.
\end{abstract}

\pacs{42.50.Gy, 42.50.Nn, 42.50.Dv, 03.67.Bg}

\maketitle

Dicke superradiance \cite{Dicke54,Eberly71,Manassah73,Agarwal74,Haroche82} remains an important problem in quantum optics primarily due to ones inability to generate entangled states of a modest number of atoms. Using single photon excitation one can produce Dicke states where only one atom out of the ensemble is excited. For this case several ground breaking experiments have been recently reported, including observation of collective Lamb shifts in regular arrays of nuclei \cite{Roehlsberger10,Scully10} or directed forward scattering from atomic ensembles in collective first excited \cite{Kimble03,Lukin03,Kimble04,Vuletic05} or Rydberg states \cite{Kuzmich12,Vuletic12,Adams12}. Beyond single-photon excited Dicke states the production of Dicke states with higher number of excitations remains a challenge. One option is the repeated measurements of photons at particular positions starting from the fully excited system.
This amounts to measuring the $m$-th order photon correlation function for $N > m$ emitters.
In this case, if the detection is unable to identify the individual photon source, the collective system cascades down the ladder of symmetric Dicke states each time a photon is recorded via projective measurements. This is another example of measurement induced entanglement among parties which do not directly interact with each other \cite{Cabrillo99,Skornia01,Kimble05,Thiel07,Monroe07,Kimble10,Weinfurter12,Hanson13}. 

The inability to distinguish the emitters is fulfilled in case of atoms confined to a region smaller than the wavelength $\lambda$ of the emitted radiation. However, if the dipole-dipole interaction between the atoms is taken into account the collective system quickly leaves the symmetric subspace populating different super- and subradiant states so that the superradiant phenomena are obscured \cite{Manassah73,Haroche82}.

The condition of indistinguishability can also be met in case of widely separated emitters as long as the detection occurs in the far field \cite{Dicke54,Eberly71,Manassah73,Agarwal74,Haroche82,Wiegner11}. 
This is fulfilled for example for atomic clouds involving many particles, relevant for most experiments in the optical domain.  
However, in this regime the superradiant characteristics depend critically on the geometry of the sample due to diffraction and propagation effects \cite{Manassah73,Haroche82} so that the superradiant behavior is concealed  by geometrical considerations.

To observe the effects of superradiance in an unobstructed manner the regime of a small number of identical widely spaced and motionless emitters appears most favorable \cite{Wiegner11}. Despite recent progress \cite{Roehlsberger10,Monroe07,Weinfurter12,Hanson13,Brewer96,Wineland05,Blatt05,Blatt11} superradiant directional spontaneous emission has not been observed for this configuration.

In what follows we focus on superradiant emission in this regime by considering a small number of identical emitters localized at positions ${\vec R}_{l}$, $l = 1, \ldots ,N$, along a linear chain with regular spacing $d \gg \lambda$ such that the dipole-dipole coupling between the emitters can be neglected (see~Fig.~1). 

\begin{figure}
\centering
\includegraphics[width=0.45\textwidth]{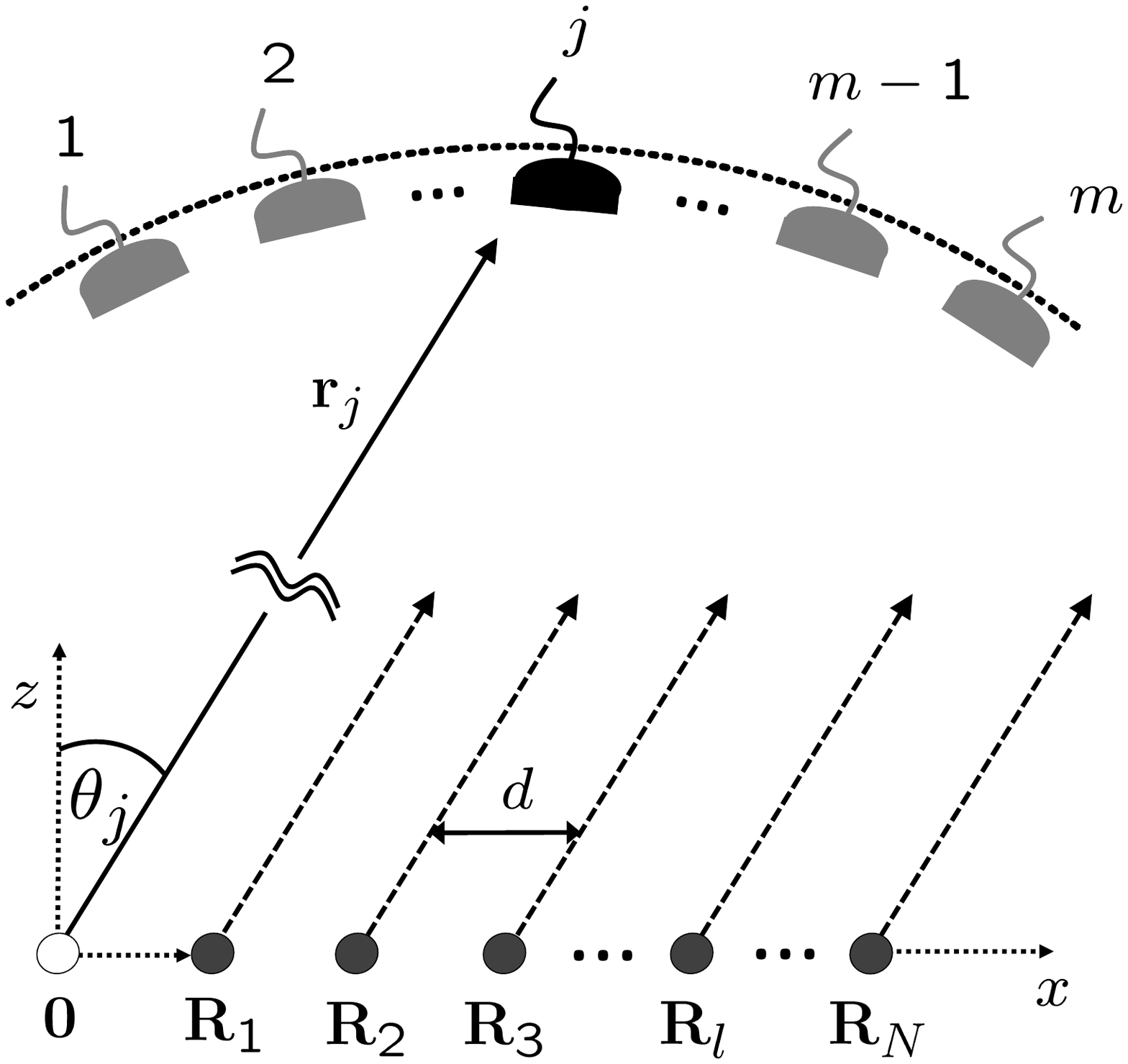}
\caption{Considered setup: $N$ identical light sources, separated by a distance $d \gg \lambda$, are placed along a chain at positions ${\vec R}_{l}$, $l = 1, \ldots , N$; the light scattered by the sources is measured by $m$ detectors, located at positions ${\vec r}_j$, $j = 1, \ldots , m$, in the far field.}
\end{figure}

We start to investigate the case of single photon emitters (SPE), e.g., $N$ two-level atoms with upper state $\ket{e_{l}}$ and ground state $\ket{g_{l}}$, $l =  1, \ldots , N$. We assume that the atomic chain is initially in the fully excited state $\ket{S_{N}}  \equiv \prod_{l = 1}^{N} \ket{e_l}$ and that $m < N$ photons spontaneously scattered by the atoms are recorded by $m$ detectors located at positions ${\vec r}_j$, $j =  1, \ldots , m$, in the far field in a circle around the sources (see~Fig.~1). For simplicity we suppose that the emitters and the detectors are in one plane and that the atomic dipole moments of the transition $\ket{e_{l}} \rightarrow \ket{g_{l}}$ are oriented perpendicular to this plane. The $m$-photon detection process can be described by the $m$-th order correlation function 
\staffeld
\label{Eq1n}
G^{(m)} (\vec{r}_1, \ldots , \vec{r}_m)  \equiv \average{:\prod_{j=1}^m \E-(\fr_j) \E+(\fr_j):} \, ,
\stoffeld
where $\average{: \ldots :}$ denotes the (normally ordered) quantum mechanical expectation value. Due to the inability to identify the particular photon sources, the electric field operator at ${\vec r}_j$ is given by $ \left[ \hat{E}^{(-)}(\vec{r}_j) \right]^\dagger = \hat{E}^{(+)}(\vec{r}_j) \sim \sum_{l=1}^{N} e^{-i\,\varphi_{lj}} \;\hat{s}^{-}_l$ \cite{Thiel07}. Here, $\hat{s}^{-}_l = |g_l\rangle\langle e_l|$ is the atomic lowering operator and $\varphi_{lj}  = - k \, \frac{\vec{r}_j\cdot\vec{R}_l}{r_j} = - l\,kd\,\sin\theta_j$ the optical phase accumulated by a photon emitted at $\vec{R}_l$ and detected at ${\vec r}_j$ relative to a photon emitted at the origin (cf.~Fig.~1).
Note that for simplicity we define the field and hence all correlation functions of $m$-th order dimensionless. The actual values can be obtained by multiplying $G^{(m)}$ with $m$ times the intensity of a single source.

Starting with all atoms in the state $\ket{S_{N}}$, we find from Eq.~(\ref{Eq1n}) for the $m$-th order correlation function, i.e., the angular distribution of the $m$-th photon after $m - 1$ photons have been recorded
\staf
\label{GMGENSa}
\begin{split}
\mathrm{G}^{(m)}_{\ket{S_{N}}}(\theta_1,...,\theta_m)  \sim ||\sum_{\substack{\sigma_1, ..., \sigma_m = 1 \\ \sigma_1 \neq ... \neq \sigma_m}}^{N} \prod_{j = 1}^{m} e^{-i\,\varphi_{\sigma_j j}} \ket{g_{\sigma_j}}||^2 \\
 = \sum_{\substack{\sigma_1, ..., \sigma_m = 1 \\ \sigma_1 < ... <  \sigma_m}}^{N} |\sum_{\substack{\sigma_1,...,\sigma_m \\\in \,{\cal S}_m}}\prod_{j = 1}^{m} e^{-i\,\varphi_{\sigma_j j}}|^2\, .
\end{split}
\stof
Here, $||\ket{\psi}||^2 = \langle \psi | \psi \rangle$ defines the norm of the state vector $\ket{\psi}$, $|...|$ abbreviate absolute values, and the expression $\sum_{\substack{\sigma_1,...,\sigma_m \\\in \,{\cal S}_m}}$ denotes the sum over the symmetric group ${\cal S}_m$ with elements $\sigma_1,...,\sigma_m$. 
In Eq.~(\ref{GMGENSa}) the products $\prod_{j = 1}^{m} e^{-i\,\varphi_{\sigma_j j}}$ represent $m$-photon quantum paths with phases $\sum_{j=1}^{m} \varphi_{\sigma_j j}$, accumulated by $m$ photons emitted from $m$ sources at $\vec{R}_{\sigma_j}$ and recorded by $m$ detectors at ${\vec r}_j$. Since 
the particular source of a recorded photon is unknown we have to sum over all possible combinations of $m$-photon quantum paths. This is expressed by the sum $\sum_{\substack{\sigma_1, ..., \sigma_m = 1}}^{N}$ in the first line of Eq. (\ref{GMGENSa}). Hereby, the condition $\sigma_1 \neq ... \neq \sigma_m$ ensures that each detector records at most one photon.
Considering that several combinations of $m$-photon quantum paths lead to the \textit{same} final atomic state and thus have to be added coherently, we end up with the modulus square in the second line of Eq.~(\ref{GMGENSa}). 
Hereby, for the $\binom{N}{m}$ \textit{different} final atomic states, the corresponding transition probabilities $|...|^2$ have to be summed incoherently, what results in the first sum $\sum_{\substack{\sigma_1, ..., \sigma_m = 1 \\ \sigma_1 < ... <  \sigma_m}}^{N}$ of the second line of Eq.~(\ref{GMGENSa}). 

We next consider that $m-1$ detectors are placed at the same position ${\vec r}_1$ and the last detector at ${\vec r}_2$. 
Under these conditions Eq.~(\ref{GMGENSa}) takes the form \cite{Oppel14}
\staf
\label{GMGENSb}
\begin{split}
& \mathrm{G}^{(m)}_{\ket{S_{N}}}(\theta_1,...,\theta_1,\theta_2) \sim \\
& \frac{N - m}{N} + \frac{m - 1}{N^2}\frac{\sin^2(N\frac{\varphi_{11}-\varphi_{12}}{2})}{\sin^2(\frac{\varphi_{11}-\varphi_{12}}{2})} \, .
\end{split}
\stof
Eq.~(\ref{GMGENSb}) corresponds to the emission pattern of a symmetric Dicke state with $N - (m - 1)$ excitations and displays the corresponding superradiant emission characteristics: even though all $N$ atoms emit spontaneously the angular distribution of the probability to detect the $m$-th photon at $\theta_2$ after $m-1$ photons have been recorded at $\theta_1$ equals the interference pattern of a coherently illuminated grating with $N$ slits, with the central maximum at $\theta_2 = \theta_1$.  
The distribution ${G}^{(m)}_{\ket{S_{N}}}(\theta_1,...,\theta_1,\theta_2)$ for the initial state $\ket{S_{N}}$, i.e., $N$ excited and spontaneously emitting atoms, is thus identical to the mean radiated intensity of a symmetric Dicke state with $N-(m-1)$ atoms in the excited state and $m-1$ atoms in the ground state.
Note that the dipole moment for any of the collective Dicke states is zero. The peaked emission pattern displayed in Eq.~(\ref{GMGENSb}) is thus not due to a synchronisation of atomic dipoles radiating in phase but due to the particular collective atom-atom correlations inherent to symmetric Dicke states \cite{Wiegner11}. 
The width $\delta\theta_2$ (FWHM) of the distribution $\mathrm{G}^{(m)}_{\ket{S_{N}}}(\theta_1,...,\theta_1,\theta_2)$ is given by
\staf
\label{peakwidth}
\delta\theta_2 \approx \frac{2\pi}{N\,k\,d}\,.
\stof
For growing numbers of emitters an increased focusing of the $m$-th photon in the direction of $\theta_1$  is thus observed. The visibility ${\cal V}_{SPE} = (m-1)/(m+1-(2m/N))$ of the distribution vanishes for $m = 1$ illustrating the fact that the atoms emit incoherently, whereas for $m = N$ a maximum visibility of ${\cal V}_{SPE} = 100 \%$ is obtained.  Fig.~2 displays $\mathrm{G}^{(m)}_{\ket{S_{N}}}(0,...,0, \theta_2)$ for $m = N$ as a function of the observation angle $\theta_2$ for $N = 2,3,5,10 $ SPE. Clearly, the width of the distribution is decreasing with increasing number of emitters $N$.

\begin{figure}[t!]
\centering
\includegraphics[width=0.48\textwidth]{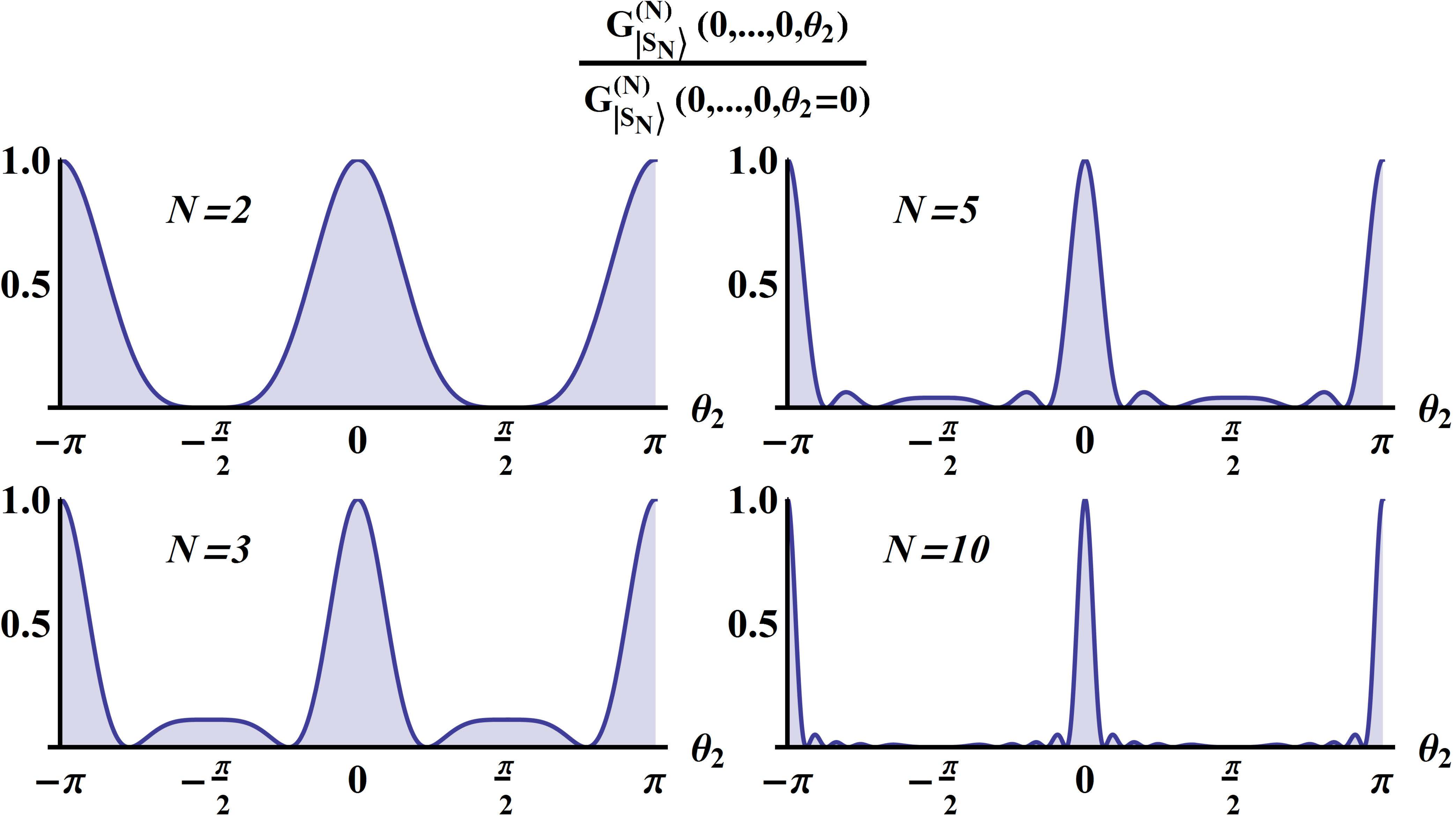}
\caption{Plot of the $m$-th order correlation function $\mathrm{G}^{(m)}_{\ket{S_{N}}}(0,...,0, \theta_2)$ for $N = m = 2,3,5,10$ single photon emitters (SPE). For a better comparison each function is normalized to its maximum value. To keep the focus to the central maximum we chose $\theta_1 = 0$ and $kd = \pi$.}
\end{figure}

Due to the particular coherence of the Dicke states an enhanced directional emission of spontaneous photons is expected to occur for correlated quantum systems only. However, the same focussed emission of incoherent photons is observed also for statistically independent incoherent classical sources. 
In this case each emitter may contribute to the $m$-th order correlation function not only a single photon but up to $m$ photons. This amounts to consider for the $m$-th order correlation function $\mathrm{G}^{(m)}_{N} (\theta_1,...,\theta_1,\theta_2)$ all possible combinations of $m_l$ photons stemming from source $l$ such that $\sum_{l=1}^N m_l = m$, or, in other words, all partitions of the number $m$, by keeping trace of the phase factors of the various $m$-photon quantum paths. The detailed calculation shows that for $N$ statistically independent incoherent classical sources each individual partition displays a focussed spatial emission pattern of the same form as given by Eq.~(\ref{GMGENSb}) \cite{Oppel14}. Superposing all partitions -- weighted with the corresponding statistics --  thus leads, apart from an offset, to the same focussed spatial emission pattern as in case of $N$ SPE. For example, for $N$ thermal light sources (TLS) with gaussian statistics we obtain \cite{Oppel14}
\staffeld
\label{GMNTLS}
\mathrm{G}^{(m)}_{N \, TLS}(\theta_1,...,\theta_1,\theta_2) \sim  1 + \frac{m-1}{N^2}\frac{\sin^2(N\frac{\varphi_{11}-\varphi_{12}}{2})}{\sin^2(\frac{\varphi_{11}-\varphi_{12}}{2})} 
\stoffeld
displaying the same probability to detect the $m$-th photon in the direction $\theta_2 = \theta_1$ after $m-1$ photons have been recorded at $\theta_1$ as in case of $N$ SPE, though with a slightly reduced visibility ${\cal V}_{TLS} = (m-1)/(m+1)$. Again, for $m = 1$ the visibility vanishes since all sources scatter incoherently, whereas for  large $m$ the visibility ${\cal V}_{TLS}$ approaches $100 \%$, independent of $N$. A similar result is obtained for $N$ statistically independent coherent light sources, with the same width $\delta\theta_2$ as in case of $N$  TLS or $N$ SPE given by Eq.~(\ref{peakwidth}) but an intermediate visibility ${\cal V}_{TLS} < {\cal V}_{CLS} < {\cal V}_{SPE}$ \cite{Oppel14}. Note that in case of classical light sources the correlation function does not vanish for $m > N$ since each light source may scatter more than one photon.

\begin{figure}[t]
\centering
\includegraphics[width=0.48\textwidth]{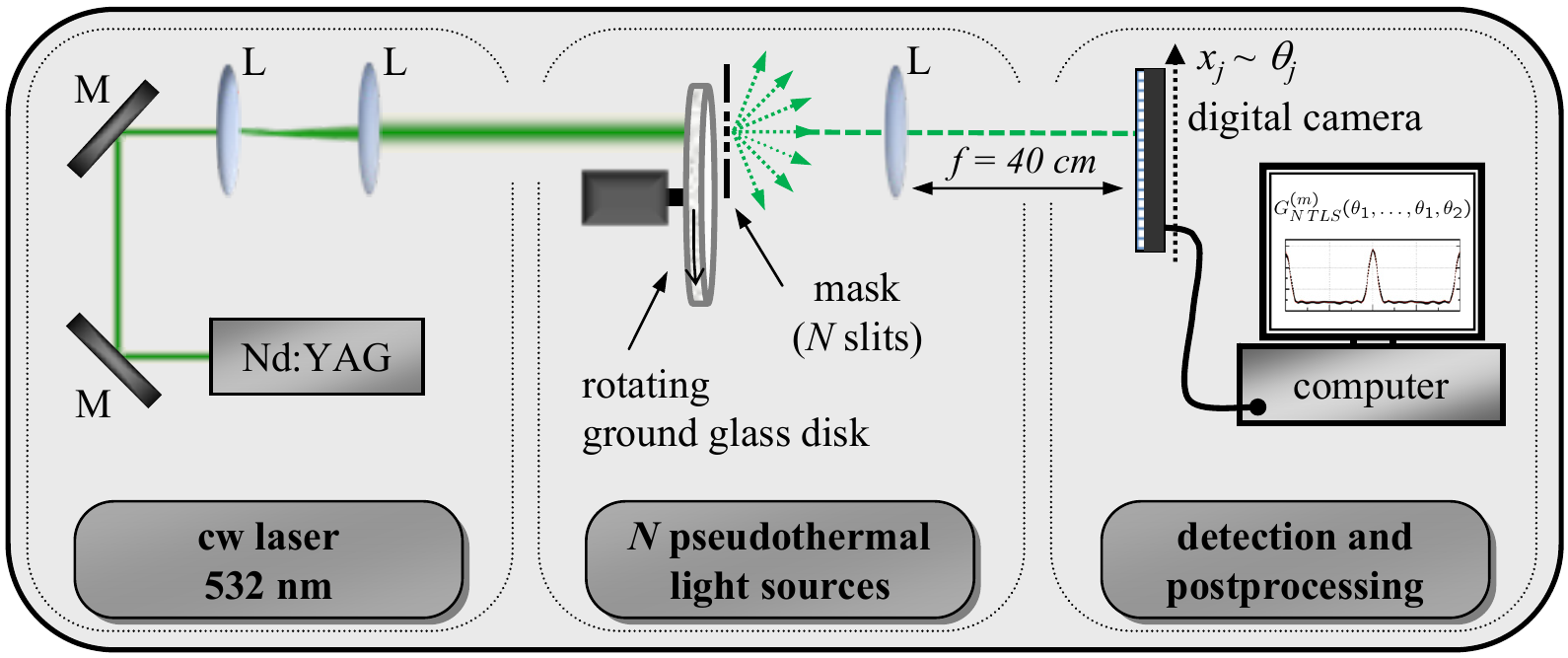}
 \caption{Experimental setup to measure $\mathrm{G}^{(m)}_{N \, TLS}(\theta_1,...,\theta_1,\theta_2)$ with $N$ pseudothermal light sources (TLS). For details see text. M: mirror, L: lens}
\end{figure}
 
To measure $\mathrm{G}^{(m)}_{N \, TLS}(\theta_1,...,\theta_1,\theta_2)$ for $N$ statistically independent incoherent TLS we used a mask with $N$ identical slits of width $a = 25 \, \mu$m and separation $d = 200 \, \mu$m, placed a few centimeters behind a rotating ground glass disk illuminated by a linearly polarized frequency-doubled Nd:YAG laser at $\lambda = 532$ nm (see Fig.~3).  
The large number of time-dependent speckles generated within each slit, produced by the stochastically interfering waves scattered from the granular surface of the ground glass disk, represent many independent point-like sub-sources equivalent to an ordinary spatially incoherent thermal source.
The coherence time of the pseudothermal sources depends on the rotational speed of the disk \cite{Tuft71} and was chosen to $\tau_c \approx 50$ ms. The incident laser beam was enlarged to 1~cm to ensure a homogeneous illumination of the mask so that all $N$ TLS radiate with equal intensity. 
Since multiphoton interferences of classical sources can be measured in the high-intensity regime \cite{Chekhova2008} we used a conventional digital camera to determine $\mathrm{G}^{(m)}_{N \, TLS}(\theta_1,...,\theta_m)$ placed in the focal point (Fourier plane) of a lens behind the mask ($z \approx f$) thus fulfilling the far field condition. Each pixel of the camera may serve as a detector to register the intensity at position $x_j/z \sim \theta_j$. 
With more than a million of pixels a digital camera has the advantage that the amount of data accumulated in one frame to correlate the intensities at $m$ different pixels is exceedingly higher than using $m$ single photon detectors \cite{Oppel12}. In order to obtain interference signals of high visibility, the integration time of the camera $\tau_{i}$ was chosen much shorter than the coherence time of the TLS, in our case $\tau_{i} \approx 1$~ms $<< \tau_{c}$. 

\begin{figure}[h!]
\centering
\includegraphics[width=0.48\textwidth]{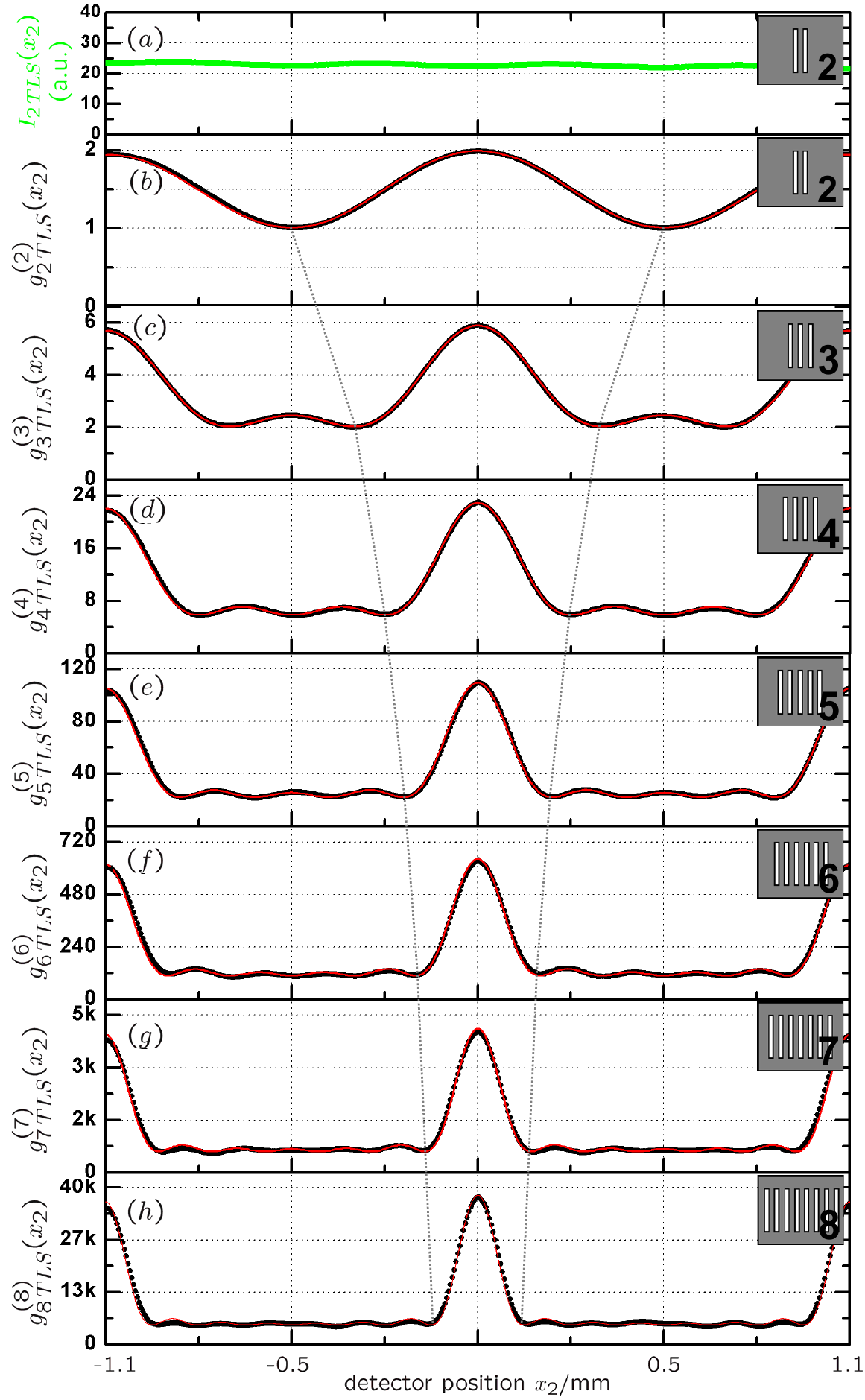}
\caption{Experimental results. (a) Average intensity $I_{2 \, TLS} (x_2)$ of 2 TLS demonstrating that the pseudothermal light sources are spatially incoherent in first order. (b)-(h) Measurement of the normalized $m$-th order correlation function $\mathrm{g}^{(m)}_{N \, TLS}(0, \ldots, 0, x_2) = \mathrm{G}^{(m)}_{N \, TLS}(0, \ldots, 0, x_2)/(I_{1 \, TLS}^{m-1} (0) I_{1 \, TLS} (x_2))$ for $m = N = 2, \ldots, 8$ as a function of the $m$-th detector at $x_2$. The focussed emission of the $m$-th photon at $x_2 = 0$ after $m - 1$ photons have been recorded at $x_1 = 0$ is clearly visible. The theoretical predictions of Eq.~(\ref{GMNTLS}) are displayed by the red curves. Fit parameters are an offset and a global prefactor.}
\end{figure}

Fig.~4 displays the experimental results for $\mathrm{G}^{(m)}_{N \, TLS}(0, \ldots, 0, x_2)$ as a function of $x_2$ for $m=N = 2, \ldots, 8$. To verify the absence of first-order coherence the averaged intensity $I_{2 \, TLS} (x_2)$ in case of two TLS was measured (see Fig. 4(a)). As expected, the intensity is constant confirming the spatial incoherence of the pseudothermal sources. The distribution $\mathrm{G}^{(m)}_{N \, TLS}(0, \ldots, 0, x_2)$ for different $m = N$ are shown in Figs. 4(b) - (h). They are in excellent agreement with the theoretical predictions of Eq.~(\ref{GMNTLS}). In particular, the increased probability to detect the $N$-th photon at $x_2 = 0$ after $N-1$ photons have been recorded at $x_1 = 0$ as a function of $N$ is clearly visible. 

The foregoing theoretical calculations and experimental results show that beyond entangled symmetric Dicke states it is also possible to employ statistically independent light sources to obtain a focussed spatial emission pattern of incoherently emitted radiation. In case of $N$ initially uncorrelated SPE, e.g., two-level atoms in the excited state, the directional spontaneous emission of the $m$-th photon is due to preceding measurements of $m-1$ photons along selected directions, projecting the uncorrelated atoms into Dicke states of excitation $N-(m-1)$ ($N \geq m > 1$). Surprisingly, the same behavior, i.e., an enhanced probability to detect the $m$-th photon at $\theta_1$ after $m-1$ photons have been recorded at $\theta_1$, is obtained also for statistically independent incoherent classical sources.  
The superradiant emission patterns generated by non-classical emitters display however a higher visibility than the one produced by the classical sources.

\textit{Acknowledgements} The authors gratefully acknowledge funding by the Erlangen Graduate School in Advanced Optical Technologies (SAOT) by the German Research Foundation (DFG) in the framework of the German excellence initiative. R.W. and S. O. gratefully acknowledge financial support by the Elite Network of Bavaria and the hospitality at the Oklahoma State University. This work was supported by the DFG research grant ZA 293/4-1.

\end{document}